\documentclass[aps,prl,amsfonts,amsmath,amssymb,reprint,twocolumn,superscriptaddress,showpacs,a4paper]{revtex4-1}

\usepackage{color}
\usepackage{graphicx}
\usepackage{hyperref}
\usepackage{siunitx}

\usepackage{bbm}

\usepackage{ulem}
\definecolor{mygreen}{rgb}{0,0.5,0} 
\definecolor{myred}{rgb}{0.75,0,0} 
\definecolor{myblue}{rgb}{0,0,0.75} 
\definecolor{mymagenta}{cmyk}{0,1,0,0.12} 
\definecolor{mycyan}{cmyk}{1,0,0,0.12} 
\definecolor{myorange}{rgb}{1,0.5,0}

\newcommand{\Vone}{\overline{V}_1}  
\newcommand{\Vtwo}{\overline{V}_2}  
\newcommand{\tgateone}{t_{1}^{({\rm gate})}}  
\newcommand{\tgatetwo}{t_{2}^{({\rm gate})}}  

\newcommand{\Ndiff}{N_{\rm diff}}  
\newcommand{\var}{{\rm var}}
\newcommand{\NL}{N_{\rm phot}}  
\newcommand{\BD}{{DPD}} 
\newcommand{\VBD}{V_{\rm \BD}}  

\newcommand{\tda}{\rm{t}^{\rm{(DA)}}}

\newcommand{\VADA}{\overline{V}_{\rm DA}}

\begin{document}

\title{
Real-time shot-noise-limited differential photodetection for atomic quantum control
}

\newcommand{\ICFOAddress}{ICFO -- Institut de Ciencies Fotoniques, Av. Carl Friedrich Gauss, 3, 08860 Castelldefels, Barcelona, Spain}

\newcommand{\ICREAAddress}{ICREA -- Instituci\'{o} Catalana de Re{c}erca i Estudis Avan\c{c}ats, 08015 Barcelona, Spain}

\author{F. Martin Ciurana}
\affiliation{\ICFOAddress}

\author{G. Colangelo}
\affiliation{\ICFOAddress}

\author{Robert J. Sewell}
\affiliation{\ICFOAddress}

\author{M.W.~Mitchell}
\affiliation{\ICFOAddress}
\affiliation{\ICREAAddress}

\date{\today}

\begin{abstract}

We demonstrate high-efficiency, shot-noise-limited differential photodetection with real-time signal conditioning, suitable for feedback-based quantum control of atomic systems. The detector system has quantum efficiency of 0.92,  is shot-noise limited from $7.4 \times 10^5$ to $3.7 \times 10^8$ photons per pulse, and provides real-time voltage-encoded output at up to $\SI{2.3}{\mega pulses per second}$. 

\end{abstract}

\maketitle

Feedback control of atomic quantum systems \cite{Serafini2012} enables quantum information protocols including deterministic teleportation \cite{sherson2006}, stabilization of non-classical states \cite{Sayrin2011}, entanglement generation \cite{TothNJP2010,BehboodPRL2014} and quantum-enhanced sensing \cite{Yonezawa2012,InouePRL2013}.  Efficient closed-loop control has been achieved by combining optical quantum non-demolition (QND) measurement \cite{SewellNP2013,hosten2016} with electromagnetic \cite{StocktonThesis2007} or optical \cite{BehboodPRL2013} feedback. Fidelity of these protocols requires speed, sensitivity and low disturbance in the QND measurement  \cite{DeutschOC2010}.  Because advanced QND techniques such as two-color probing \cite{AppelPNAS2009} and two-polarization probing \cite{KoschorreckPRL2010a} address specific hyperfine transitions, this also requires small ($\sim$ GHz) detunings, and thus low photon numbers to achieve low disturbance.  This combination of requirements in the measurement places multiple demands on the detectors used in the QND measurement.  

Here we present a balanced differential photodetector (\BD) suitable for $\sim$ \SI{}{\micro \second} pulsed Faraday rotation \cite{TakanoPRL2009} and two-color \cite{AppelPNAS2009} dispersive probing of atomic ensembles with sensitivity and dynamic range comparable to the best published differential detectors \cite{HansenOL2001,WindpassingerMST2009}. The {\BD} employs $> 90$\% quantum efficiency photodiodes and charged-particle-detection amplifiers. Real time output is achieved with low-noise sample and hold amplifiers (SHAs) and a differential amplifier (DA). The system enables atomic quantum control in which feedback to an atomic system must be accomplished with a sub-\SI{}{\micro s} loop time \cite{MabuchiIJRNC2005,BehboodPRL2013}. 

\begin{figure}[h]
       \centering     
             \includegraphics[width=0.9\columnwidth]{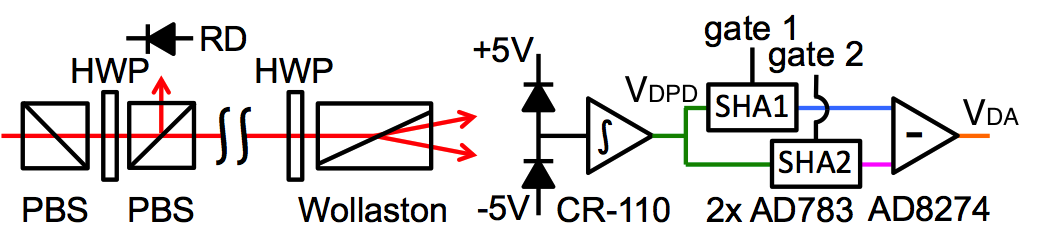} \\
                 \includegraphics[width=0.9\columnwidth]{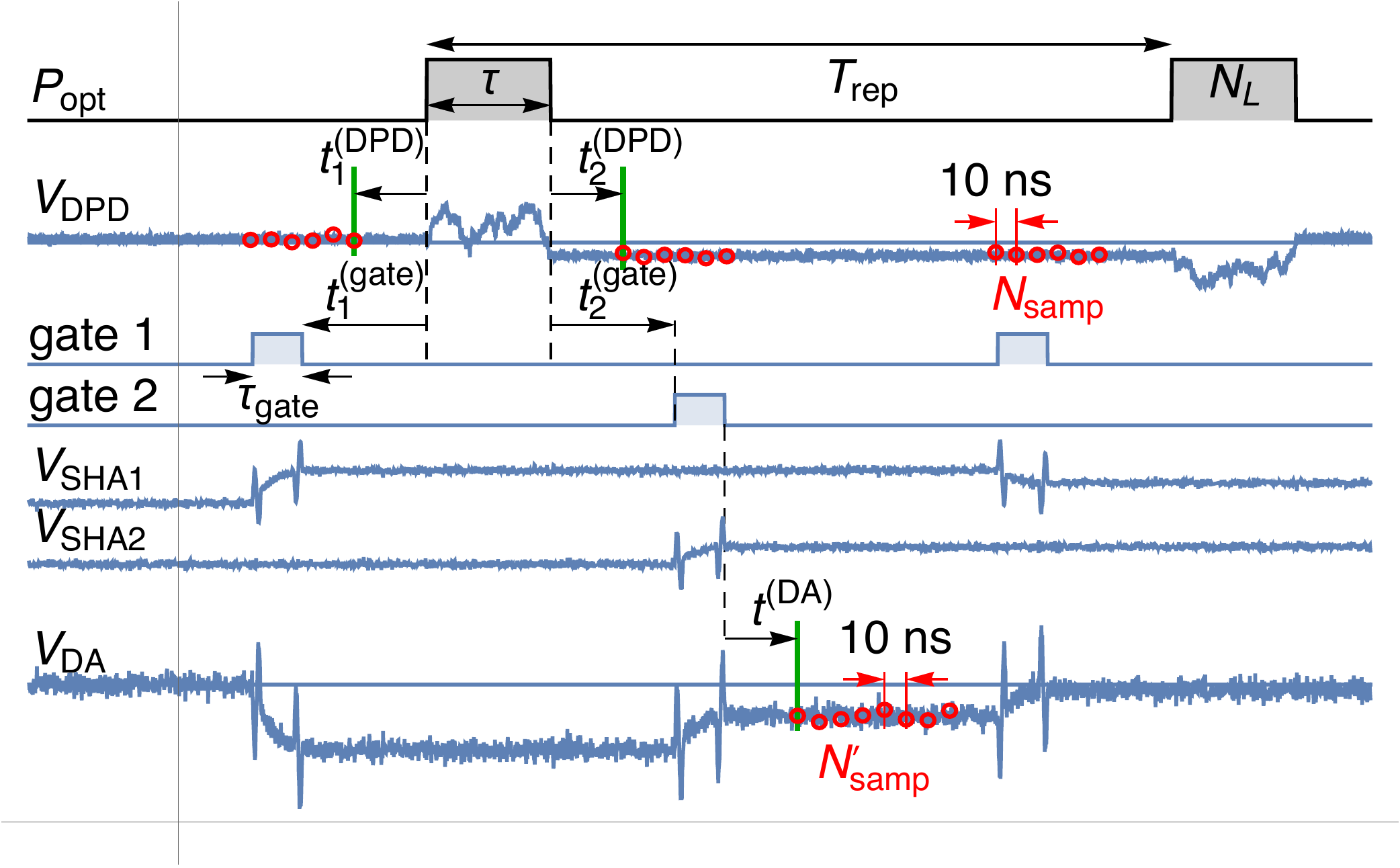}
                                                     
              \caption{Top: Schematic of the optics and detector electronics. A laser and acousto-optic modulator (not shown) are used to produce pulses of desired duration and energy. A first polarizing beam splitter (PBS) is used to generate a well defined linear polarization, a half-wave plate (HWP) and second PBS are used to split a constant fraction of the pulse to a reference detector (RD).  A HWP together with a Wollaston prism are used to balance the energies reaching the photodiodes, wired in series for direct current subtraction.  A charge-sensitive pre-amplifier (Cremat CR-110) amplifies the difference current, and a pair of SHAs (Analog Devices  AD783) sample the pre-amplifier output shortly before and shortly after the pulse.  A differential amplifier (DA)  (Analog Devices AD8274) outputs the difference of the two SHA signals.
              Bottom: Timing diagram of optical input, SHA gate voltages and {\BD } and DA signals, illustrating a possible response to two pulses of a pulse train. $P_{\rm opt}$: optical power, $V_{\rm \BD}$: Balanced detector output, gate 1, gate 2: gate voltages causing the respective SHAs to sample (high) and to hold (low),  $V_{\rm DA}$: differential amplifier output.  Red circles show oscilloscope voltage samples use to characterize the {\BD } and DA noise characteristics. }
                \label{fig:Setup}  
 \end{figure}

The electronics for the {\BD} and DA together with the test setup are shown schematically in Fig.~\ref{fig:Setup}. The detector consists of two PIN photo-diodes (PDs) (Hamamatsu S3883)  connected in series and reverse biased by \SI{5}{V} to improve their response time.  The differential output current is DC coupled to the integrator, a very low noise charge-sensitive pre-amplifier (Cremat CR-110) with a capacitor $C_{\rm i}$ and a discharge resistor $R_{\rm i}$ in the feedback branch. Together these determine the relaxation time constant $\tau= R_{\rm i} C_{\rm i} = \SI{290}{\micro s}$ of the integrator. The \SI{50}{ns} rise-time of the circuit is limited by the capacitance of the photodiodes while the CR-100 itself has a nominal rise time of \SI{7}{ns}.  
Pulses longer than the rise time but shorter than the relaxation produce a step in the output voltage proportional to $\Ndiff$, the difference of photon numbers on the two photodiodes.  
The output from the {\BD} is captured by a pair of sample and hold amplifiers (SHAs) (Analog Devices AD783), gated with TTL signals. The SHA1 captures the voltage of the {\BD} before the optical pulse arrives, and SHA2 captures it after the end of the pulse. A differential amplifier (DA) (Analog Devices AD8274) amplifies the difference of the two voltages held on the SHAs.

To characterize the noise performance of the {\BD} we send trains of pulses with a desired photon number $N$, pulse duration  $\tau$, and pulse repetition period $T_{\rm rep}$. We use a continuous-wave diode laser at \SI{780}{nm}, chopped by an acousto-optic modulator and balanced by means of a half waveplate and Wollaston prism, as shown in Fig. ~\ref{fig:Setup}.  We record the reference detector (RD) and {\BD} output voltages on an 8-bit digital storage oscilloscope (LeCroy Waverunner 64Xi) at a sampling rate of 100 Msps which acquires samples continuously and asynchronously to the pulse generation. 
We define a single measurement for the {\BD} as $\Ndiff = C(\Vone - \Vtwo)$, where $\overline{V}_{1}(\overline{V}_{2})$ is the mean of $N_{\rm{samp}}$ voltage samples before (after) the optical pulse and $C$ is a calibration factor. The  number of photons  $N_{\rm phot}$ in a pulse is estimated as $N_{\rm phot} = C_{\rm RD} \sum_i {V}_{\rm RD}(t_i)$ where  $V_{\rm RD}$ is the voltage output of RD. The sum is taken over the duration of the pulse and $C_{\rm RD}$ is a calibration factor obtained by comparison against a power meter.  For a given set of conditions, we adjust the waveplate to give a balanced signal $\Ndiff \approx 0$ and record $M$ pulses in a single pulse train, from which we extract $M$ values for $\Ndiff$ and $N_{\rm phot}$ and compute statistics.

\newcommand{\Trep}{T_{\rm rep}}
\newcommand{\Nsamp}{N_{\rm samp}}

When source and detector fluctuations are taken into consideration, a linear detector will have an output signal variance given by a second-order polynomial in the average optical input energy \cite{bachor2004Book}, 
\begin{equation}
\var \Ndiff = a_0 \NL^0 + \eta \NL + a_2\NL^2.
\label{eq:Polynomial}
\end{equation}
Here $a_0$ is the ``electronic noise'' (EN) contribution, $a_2 \NL^2$ is the ``technical noise'' (TN) and the second term is the shot noise (SN) contribution with $\eta$ the quantum efficiency of the detector.  The different scalings with $\NL$ allow an unambiguous identification of the different noise contributions. 

\begin{figure}[h!]
        \centering     
                 \includegraphics[width=0.90\columnwidth]{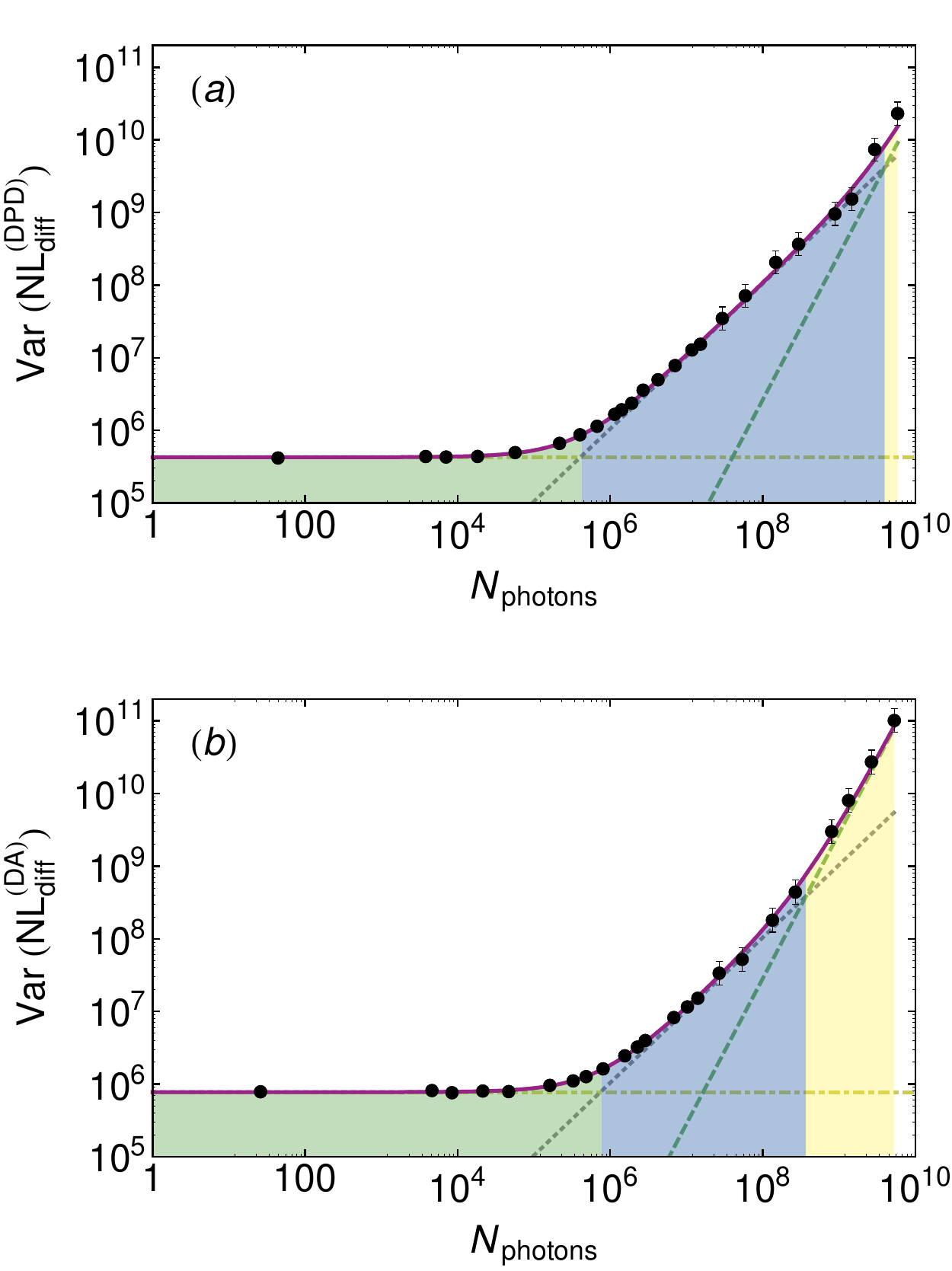}
                         \caption{Variance of the output signal of the {\BD} (a) and DA (b) as function of the input photon-number in log-log scale. Solid red line fit to ${\rm Var}(\textrm{NL}_\textrm{{diff}})$ using expression Eq.1. Shaded areas depicts the different detection responses: green EN-limited, blue SN-limited and yellow TN-limited. Error bars represent $\pm 1 \sigma$ standard error. For both devices $\Trep = \SI{0.8}{\micro s}$ and $\tau=\SI{200}{\nano s}$. 
        (a): $\var \Ndiff^{\rm{\BD}}$ as function of the ${\NL}$. Analysis done with $\textrm{t}_1^\textrm{(DPD)}=\SI{10}{\nano s}$, $\textrm{t}_2^\textrm{(DPD)}=\SI{90}{\nano s}$ and ${\Nsamp}=10$ points.  The yellow line dot-dashed is the electronic noise level, $a_0=(4.26\pm0.05)\times 10^5$, the dotted gray line is the shot-noise term, $\eta=1.05\pm0.01$, and the dashed green line is the technical noise contribution,  $a_2=(2.64\pm1.45)\times 10^{-10}$.
        (b): $\var \Ndiff^{\rm{DA}}$  as function of the ${\NL}$. The timings for the gates of the SHAs are $\tgateone =\SI{10}{\nano s}$, $\tgatetwo =\SI{20}{\nano s}$ with analysis parameters ${\tda}=\SI{170}{\nano s}$ and $N'_{\rm{samp}} =10$ points. The yellow line dot-dashed is the EN level, $a_0=(7.75\pm0.09)\times 10^5$, the dotted gray line is the SN term, $\eta=1.04\pm 0.01$, and the dashed green line is the TN contribution, $a_2=(2.84\pm0.70)\times 10^{-9}$.
        }
                 \label{fig:VarNL}  
 \end{figure}

To estimate the coefficients of Eq.1 we collect data at a variety of $\NL$ in each case recording a train of 2500 pulses with repetition period $\Trep = \SI{0.8}{\micro s}$ and pulse duration  $\tau=\SI{200}{\nano s}$. Our light source is not powerful enough to measure the turning point from SN-limited to TN-limited. In order to estimate when the TN becomes the dominant source of noise we construct ``composite pulses'' containing a larger total number of photons by summing the signals from multiple pulses \cite{Koschorreck2010PRL}.
The measured variances are fitted with Eq.~(\ref{eq:Polynomial}) to obtain $a_0$, $\eta$ and $a_2$. 
We set $\textrm{t}_1^\textrm{(DPD)}=\SI{10}{\nano s}$ to ensure that $\Vone$ is measuring the voltage before the detection of the optical pulse and $\textrm{t}_2^\textrm{(DPD)} = \SI{90}{\nano s}$ which is the minimum time to sample $> 99\%$ of the {\BD} signal, with $\Nsamp=10$ points. Typical results are shown in Fig. ~\ref{fig:VarNL} (a). 
The detector is shot-noise limited when $ a_0/\eta<\NL <\eta/a_2$. From the fit outputs, see Fig.~\ref{fig:VarNL} for details, we determine that the {\BD} is SN-limited from $(4.06\pm 0.07)\times 10^5<\NL <  (3.97 \pm 2.18)\times 10^9$ photons, i.e., its SN limited behavior extends over 4 orders of magnitude.

To characterize the noise properties of the DA we repeat the same procedure as for the {\BD}: we record on the oscilloscope the RD and DA output voltages. The SHA1 captures $\VBD(\tgateone)$ before the optical pulse arrives, and the SHA2 captures $\VBD(\tgatetwo)$ after the end of the pulse, analogous to $\Vone$ and $\Vtwo$, respectively. We define a single measurement as ${N_{\rm{diff}}}=C'{\VADA}$, where ${\VADA}$ is the mean of $N'_{\rm{samp}}$ voltage samples a time ${\tda}$ after the end of the SHA2 and $C'$ is a calibration factor obtained by comparison against a power meter.
Under the same experimental conditions, $\Trep = \SI{0.8}{\micro s}$ and $\tau=\SI{200}{\nano s}$, we record a train of 2500 pulses for each value of $\NL$ and fit $\var \Ndiff^{\rm{(DA)}}$ with Eq.~(\ref{eq:Polynomial}) we obtain $a_0$, $\eta$ and $a_2$. As before, we construct ``composite pulses" to determine $a_2$. The SHAs are gated for $\tau_{\rm{gate}}=\SI{100}{ns}$ at times $\tgateone =\SI{10}{\nano s}$ and $\tgatetwo =\SI{20}{\nano s}$. The analysis parameters are ${\tda}=\SI{170}{\nano s}$ and $N'_{\rm{samp}}=10$ points. 
Typical results are shown in Fig.~\ref{fig:VarNL} (b). From the fit outputs we determine that the DA is shot-noise limited from $(7.43\pm 0.14)\times 10^5<\NL < (3.67\pm 0.91)\times 10^8$ photons, i.e., over almost 3 orders of magnitude.

From the coefficients $a_0$ we can deduce the noise-equivalent charge (NEC), the number of photo-electrons necessary to create a signal equivalent to the electronic noise 
$q_{\rm{SN}}=\eta_{Q}\sqrt{N_{\rm{phot, ~SN}}}$ for the {\BD} and for the DA. The quantum efficiency of the photo-diodes in the detector at 780nm is $\eta_Q$= 0.92, resulting in NEC$^{\rm{\BD}}$= 600 electrons and NEC$^{\rm{DA}}$= 808 electrons. Since the two calibration experiments  were taken under the same experimental conditions, i.e., $\Trep = \SI{0.8}{\micro s}$ and $\tau=\SI{200}{\nano s}$, and comparable analysis conditions $\tau_{\rm{gate}}=\SI{100}{\nano s}$ and ${\Nsamp}=N'_{\rm{samp}}=10$ points, we see that the capability of having the signal available in real time has the cost of increasing the electronic noise level by 1.3dB.

The electronic noise of {\BD } contains high-bandwidth noise, e.g. Johnson noise, that can be reduced by averaging the in-principle constant output over a time window, which could be longer than the pulse itself.  On the other hand, longer windows will be more sensitive to drifts and ``$1/f$'' noise.  We investigate this trade-off by changing $\Nsamp$ used to obtain $\Vone$ and $\Vtwo$ and then fit $\var \Ndiff^{\rm{(DPD)}}$ with Eq.1 to get $a_0$, $\eta$ and $a_2$. The experiment is done at $\Trep = \SI{30}{\micro s}$ and $\tau=\SI{200}{\nano s}$, and the analysis with $\textrm{t}_1^\textrm{(DPD)}=\SI{10}{\nano s}$ and $\textrm{t}_2^\textrm{(DPD)}=\SI{90}{\nano s}$. For each $\NL$ we record more than 300 pulses in a single pulse train.  From these parameters we evaluate the SN limited region of the {\BD} as a function of the measurement bandwidth. We fit the EN (TN) limited region $a_0/\eta<\NL$ ($\NL>\eta/a_2$) with the polynomial $\alpha_1\Nsamp^{\beta_1} +\alpha_2 \Nsamp^ {\beta_2}$ (Eq.2), where the two terms are for the two noise time-scales. 

\begin{figure}[htp]
                \includegraphics[width=0.9\columnwidth]{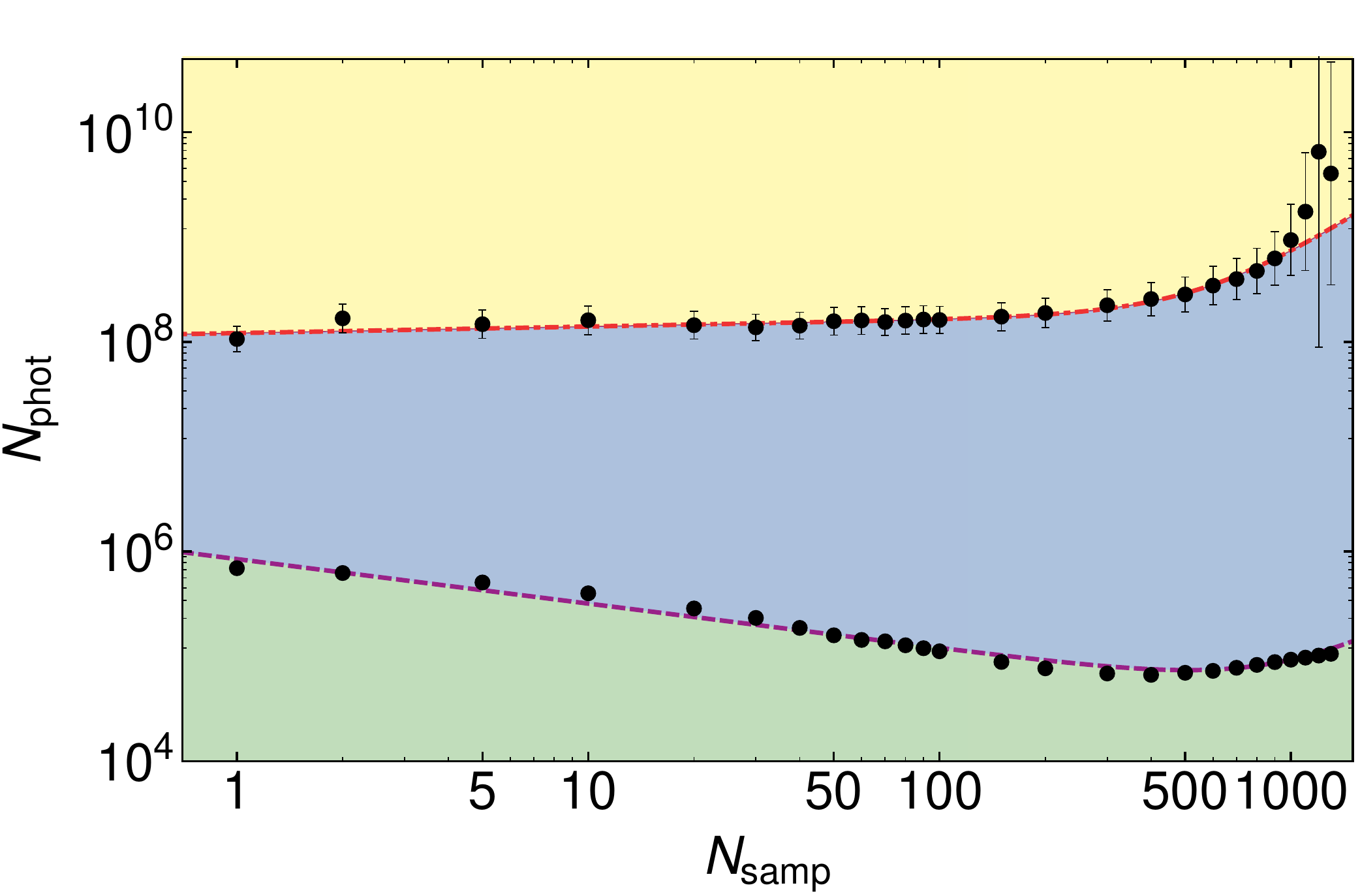}
               \caption{  {\BD } SN-limited region (blue area) as function of measurement bandwidth ($N_{\rm{samp}}$) in log-log plot. EN-limited region (green) and TN-limited (yellow), see text for details. Experimental parameters $\Trep = \SI{30}{\micro s}$ and $\tau=\SI{200}{\nano s}$. Red dot-dashed curve is a fit of TN with Eq. 2 with results $\alpha_1^{\rm{TN}}= (1.36\pm0.56)\times 10^8$, $\alpha_2^{\rm{TN}}= (0.53\pm4.65)\times 10^3$, $\beta_1^{\rm{TN}}= 0.03\pm0.11$ and $\beta_2^{\rm{TN}}= 1.99\pm1.31$. Purple dashed curve is a fit of EN with Eq.2 with output values $\alpha_1^{\rm{EN}}= (1.68\pm0.11)\times 10^6$, $\alpha_2^{\rm{EN}}= 40\pm25$, $\beta_1^{\rm{EN}}= -0.60\pm0.02$ and $\beta_2^{\rm{EN}}= 1.08\pm0.09$. The scope ``electronic noise" $\textrm{a}_0$ fit (not shown) has values $\alpha_1^{\rm{scope}}=(3.79\pm0.47)\times 10^4$, $\alpha_2^{\rm{scope}}= 0.17\pm 0.16$, $\beta_1^{\rm{scope}}= -0.96\pm0.03$ and $\beta_2^{\rm{scope}}= 0.99\pm0.13$ and conclude that the scope noise contribution is negligible. Sample rate is 100 Msps or $\SI{10}{\nano s/sample}$. Error bars represent $\pm 1 \sigma$ standard error.  }
              
                            \label{fig:VarENTimeV}
\end{figure}

In Fig.~\ref{fig:VarENTimeV} we observe a transition from $\rm{EN}$ $\propto \Nsamp ^{\beta_1}$, where $\beta_1 <0$ for $\Nsamp \lesssim 100$ points, describing the effects of averaging, to a $1/f$ regime for $\Nsamp \gtrsim 600$ points, with $\rm{EN}$ $\propto  \Nsamp^{\beta_2}$ where $\beta_2 >0$. The fit results are $\beta_1^{{\rm{EN~\BD}}}=-0.60\pm0.02$ and ${\beta_2^{{\rm{EN~\BD}}}}=0.99\pm0.13$. We also notice that Fig.~\ref{fig:VarENTimeV} shows that increasing ${\Nsamp}$ from $1$ point to $400$ we can reduce the electronic noise of the {\BD} by \SI{10.2}{dB}, and that at 400 samples the {\BD} electronic noise is minimal with a NEC of 242 electrons corresponding to a measurement bandwidth of $\SI{125}{kHz}$.

We repeat the measurement under the same scope settings to determine the electronic noise contribution of the scope itself terminating it with a 50 $\Omega$ terminator. Analogously, we vary $\Nsamp$ to obtain $\Vone$ and $\Vtwo$ and observe that the $a_0^{\rm{scope}}$ is negligible relative to $a_0^{\rm{DPD}}$. The fact that ${\beta_1^{\rm{EN~{\BD}}}}=-0.60$ and not $-1$ (as in the case of ${\beta_1^{\rm{scope}}}$) means that there is some correlated noise in the {\BD} output signal. This is expected as the \SI{100}{MHz} sampling frequency exceeds the oscilloscope input bandwidth at this setting. The measured \SI{-3}{dB} oscilloscope bandwidth is \SI{30}{MHz}.

Even though the EN increases for $\Nsamp>400$, the SN limited region i.e., the area between the EN and the TN curves, still increases with $\Nsamp$ as the reduction of the TN-limited region compensates the increase of the EN. We can observe that the TN is almost flat for $\Nsamp\lesssim 300$ but rapidly decreases for $\Nsamp > 500$. The {\BD} is SN-limited over measurements bandwidth running from $\SI{3}{MHz}$ to $\SI{35}{kHz}$.

The {\BD} presented here offers a significant improvement in speed compared to other state-of-the-art detectors, while also having somewhat lower noise \cite{HansenOL2001,WindpassingerMST2009,Takeuchi2006}. In \cite{WindpassingerMST2009} two detectors based on two different charge-sensitive pre-amplifier are described with minimal ENC=280 (Amptek-based detector) and ENC=340 (Cremat-based detector) operated at speeds $\lesssim \SI{200}{kHz}$ (exact value not reported). Our {\BD} shows a minimal ENC=242 at $\SI{125}{kHz}$, representing a noise improvement of 0.63dB (Amptek-based detector) and 1.84dB (Cremat-based detector) while operating at similar measurement bandwidth. Similarly, in \cite{Takeuchi2006} the maximum measurement speed is $\SI{200}{kHz}$ and SN limited starting from $10^6$ photons/pulse (our {\BD} is SN limited from $7\times10^4$). Ref \cite{HansenOL2001}, working at a repetition rate of 1 MHz reports a NEC of 730 electrons, whereas our {\BD} has a NEC of 600 electrons at $\SI{1}{MHz}$, a reduction of 0.87 dB. Furthermore, our {\BD} has a minimum repetition period of $T_{\rm rep}^{(\rm \BD)}$ = {\BD} rise time ($\SI{50}{\nano s}$) +  $\textrm{t}_1^\textrm{(DPD)}$ ($\SI{10}{\nano s}$) + $\textrm{t}_2^\textrm{(DPD)}$ ($\SI{10}{\nano s}$) + 2 $\Nsamp$ ( $2\times \SI{10}{\nano s}$) = $\SI{90}{\nano s}$, or equivalently, a maximum detection bandwidth of $\SI{11}{\mega Hz}$, which to our knowledge makes it the fastest quantum-noise limited differential photodetector for this energy regime, i.e., for pulses with as  few as $6.8\times 10^5$ photons. Along with the speed, our {\BD} is SN limited over measurements bandwidth running from $\sim\SI{10}{MHz}$ to kHz.

Atomic experiments have coherence times running from $\mu$s to seconds requiring a real time detector with low latency and large bandwidth to perform many manipulations of the atomic state before decoherence occur. The maximum measurement speed of our DA is determined by $\tau_{\rm{gate}}$, the sampling time of the SHAs to faithfully capture $V_{\rm \BD}$, and ${\tda}$, the settling time of the SHA once the sampling has been done. 

We investigate the effect of $\tau_{\rm{gate}}$ by measuring $\var \Ndiff^{\rm{DA}}$ vs ${\NL}$ for different values of $\tau_{\rm{gate}}$ and compare the fit outputs. We obtain the same results for $\tau_{\rm{gate}}=\SI{250}{\nano s}$, the manufacturer recommended value, as for $\tau_{\rm{gate}}=\SI{100}{\nano s}$, with fit parameters comparable within the standard error, but not for $\SI{50}{\nano s}$ where the DA-output is independent of $\NL$, i.e., dominated by EN. 

\begin{figure}[h]
        \centering     
              \includegraphics[width=0.9\columnwidth]{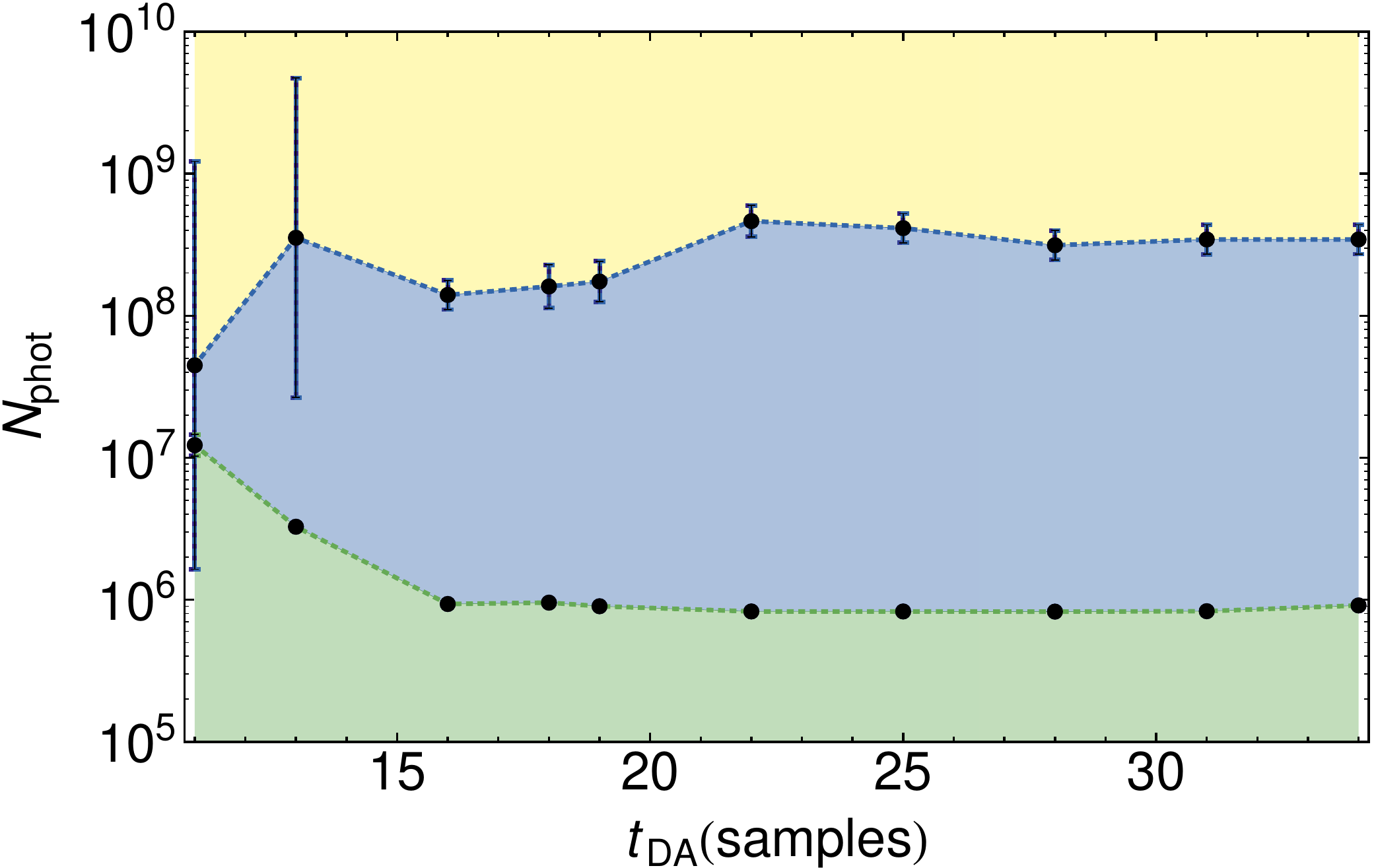}
               \caption{ DA SN-limited region (blue area) as function of the SHA2 settling time ($\tda$). Green shaded area is EN-limited region and TN-limited region in yellow, see text for details. Experimental parameters $\Trep = \SI{0.8}{\micro s}$, $\tau=\SI{200}{\nano s}$, $\tgateone=\SI{10}{\nano s}$ and $\tgatetwo=\SI{20}{\nano s}$. Analysis done with $N'_{\rm{samp}} =1$ point at a sample rate of 100 Msps or $\SI{10}{\nano s/sample}$. Error bars represent $\pm 1 \sigma$ standard error.  
              }
               \label{fig:TDA}  
 \end{figure}

We study the effect of the settling time of the SHA2 by varying the samples used for ${\VADA}$ with ${\tda}$ and fit $\var \Ndiff^{\rm{(DA)}}$ to obtain the parameters $a_0$, $\eta$ and $a_2$.  From the fit outputs we determine the EN-limited and TN-limited regions. In Fig.~\ref{fig:TDA} we see that for values of ${\tda}$ where the noise of the SHA has not had time to settle the EN and the TN contributions are large, dominant over SN. We also observe that once ${\tda}$ is sufficient, the EN region is flat as expected from the output of a the DA. From the fit outputs $\eta$ we determine that the minimum value to have $>99\%$ of the signal corresponds to  ${\tda}=\SI{170}{\nano s}$. 

The minimum time at which we can measure two consecutive pulses is given by $T_{\rm rep}^{(\rm DA)}$ = {\BD} rise time ($\SI{50}{\nano s}$) + $2 \tau_{\rm{gate}}$ ($2\times \SI{100}{\nano s}$) + ${\tda}$ 
($\SI{170}{\nano s}$) + $N'_{\rm{samp}}$ ($\SI{10}{\nano s}$)= $\SI{430}{\nano s}$, or equivalently, the maximum detection bandwidth in real time is $\SI{2.3}{\mega pulses/s}$.

Comparing Fig.~\ref{fig:VarNL} and Fig.~\ref{fig:TDA} we see that the SN-limited region is a bit narrower in Fig.~\ref{fig:TDA}, due to the different $N'_{\rm{samp}}$ used in the analysis. This suggests that the output of the DA has fast frequency noise components that could be filtered to obtain the same noise performance as in Fig.~\ref{fig:VarNL}. 

In conclusion, we have demonstrated a pulsed differential photodetector (\BD) and a detection system to make the signal available in real time. The {\BD} has bandwidth up to $\sim \SI{11}{\mega Hz}$ which to our knowledge makes it the fastest quantum-noise limited differential photodetector for pulses with as  few as $6.8\times 10^5$ photons per pulse. We make the signal available in real time by using a pair of sample and hold amplifiers (SHA) and a differential amplifier (DA). The DA is shot noise limited per input pulses varying from $7.4\times10^5$ to $3.7\times 10^8$ photons per pulse and shows low latency, $\SI{170}{\nano s}$. The {\BD} together with the DA and can directly be employed in real time quantum control experiments with flexible measurement bandwidth varying from $\SI{}{\kilo Hz}$ up to $\SI{2.3}{\mega Hz}$. 

We thank Patrick Windpassinger for help and insights in the detector design, and Jos\'{e} Carlos Cifuentes for help with the electronics design and construction.

Work supported by the Spanish MINECO projects MAQRO (Ref. FIS2015-68039-P), EPEC (FIS2014-62181-EXP) and Severo Ochoa grant SEV-2015-0522, Catalan 2014-SGR-1295, by the European Research Council project AQUMET, Horizon 2020 FET Proactive project QUIC and by Fundaci\'{o} Privada CELLEX.

\bibliographystyle{apsrev}
\bibliography{FastCrematBib2}

\end{document}